\documentclass[conference]{IEEEtran}
\IEEEoverridecommandlockouts
\usepackage{cite}
\usepackage{amsmath,amssymb,amsfonts}
\usepackage{algorithmic}
\usepackage[pdftex]{graphicx}
\graphicspath{{../figures/}}
\usepackage{lipsum}
\ifCLASSOPTIONcompsoc
\usepackage[caption=false, font=normalsize, labelfont=sf, textfont=sf]{subfig}
\else
\usepackage[caption=false, font=footnotesize]{subfig}
\fi
\usepackage{booktabs,caption}
\usepackage[flushleft]{threeparttable}

\usepackage{textcomp}
\usepackage{xcolor}

\DeclareGraphicsExtensions{.pdf,.jpeg,.png}

\usepackage{mathabx}

\usepackage{array}
\usepackage{mdwmath}
\usepackage{mdwtab}
\usepackage{eqparbox}
\usepackage{url}
\usepackage{comment}
\def\BibTeX{{\rm B\kern-.05em{\sc i\kern-.025em b}\kern-.08em
    T\kern-.1667em\lower.7ex\hbox{E}\kern-.125emX}}
\begin{document}

\title{A system design approach toward integrated cryogenic quantum control systems\\
}

\author{Mridula Prathapan, Peter Mueller, David Heim, Maria Vittoria Oropallo, Matthias Br\"{a}ndli,\\ Pier Andrea Francese, Marcel Kossel, Andrea Ruffino, Cezar Zota, Eunjung Cha, and Thomas Morf \\


\IEEEauthorblockA{\textit{IBM Zurich Research Laboratory, R\"{u}schlikon, Switzerland.} Email: mrp@zurich.ibm.com}
}
\maketitle

\begin{abstract}
In this paper, we provide a system level perspective on the design of control electronics for large scale quantum systems. Quantum computing systems with high-fidelity control and readout, coherent coupling, calibrated gates, and re-configurable circuits with low error rates are expected to have superior quantum volumes. Cryogenic CMOS plays a crucial role in the realization of scalable quantum computers, by minimizing the feature size, lowering the cost, power consumption, and implementing low latency error correction. Our approach toward achieving scalable feed-back based control systems includes the design of memory based arbitrary waveform generators (AWG's), wide band radio frequency analog to digital converters, integrated amplifier chain, and state discriminators that can be synchronized with gate sequences. Digitally assisted designs, when implemented in an advanced CMOS node such as 7 nm can reap the benefits of low power due to scaling. A qubit readout chain demands several amplification stages before the digitizer. We propose the co-integration of our in-house developed InP HEMT LNAs with CMOS LNA stages to achieve the required gain at the digitizer input with minimal area. Our approach using high impedance matching between the HEMT LNA and the cryogenic CMOS receiver can relax the design constraints of an inverter-based CMOS LNA, paving the way toward a fully integrated qubit readout chain. 
The qubit state discriminator consists of a digital signal processor that computes the qubit state from the digitizer output and a  pre-determined threshold. The proposed system realizes feedback-based optimal control for error mitigation and reduction of the required data rate through the serial interface to room temperature electronics.
\end{abstract}

\begin{IEEEkeywords}
Quantum control electronics, cryogenic CMOS, spin qubit systems, optimal control, error mitigation, quantum error correction, qubit state detector, integrated readout chain. 
\end{IEEEkeywords}

\section{Introduction}
The architecture of any solid state-based quantum computation scheme, 
requires a platform of coherent physical qubits, strong tunable coupling with high fidelity control and readout system, together with error correction schemes to realize fault tolerant logical operations \mbox{\cite{b1}-\cite{b2}}. The quantum computing stack consists of different layers, that can be broadly classified into quantum hardware layer, quantum control layer and the application layer. Quantum hardware layer contains the qubits and its ancillary elements to realize quantum gates. The control layer is responsible for the control and readout of quantum gates, by facilitating the integration and scalability of the overall system. The control layer also contains gate sequencing and error correction logic, hosted on a general-purpose processor at room temperature, whose main function is to sequence and orchestrate the execution of quantum algorithms. The applications layer translates real-world problems into quantum algorithms, hosted on a server or may use the quantum serverless approach from IBM \cite{b3}. 

\begin{figure*}[tbh]
	\centering
	\subfloat[Proposed scalable logical qubit control system\label{1a}]{%
	\includegraphics[width=0.33\linewidth]{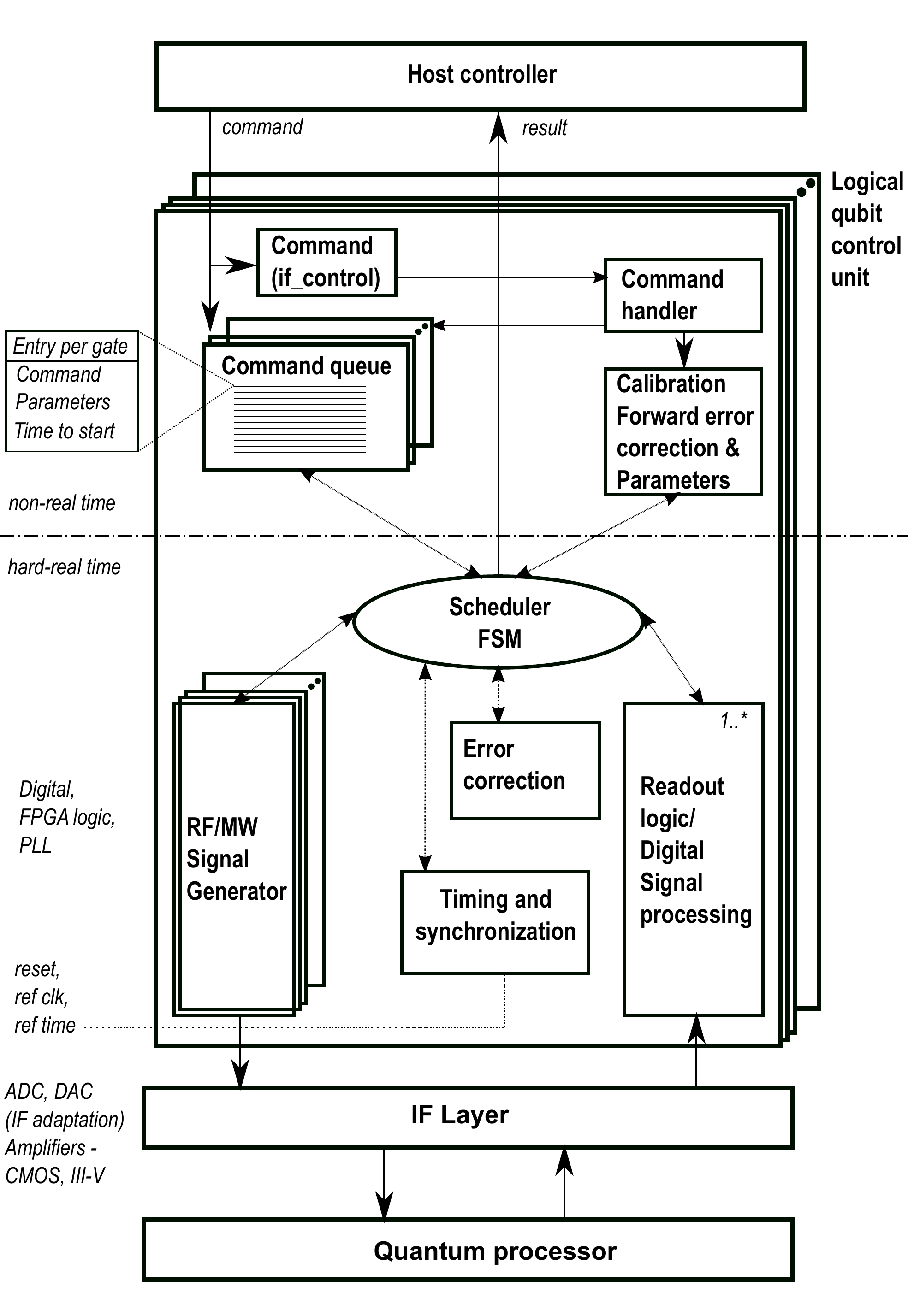}}
\hfill
\subfloat[Proposed integrated cryogenic control system for spin qubits \label{1b}]{%
	\includegraphics[width=0.33\linewidth]{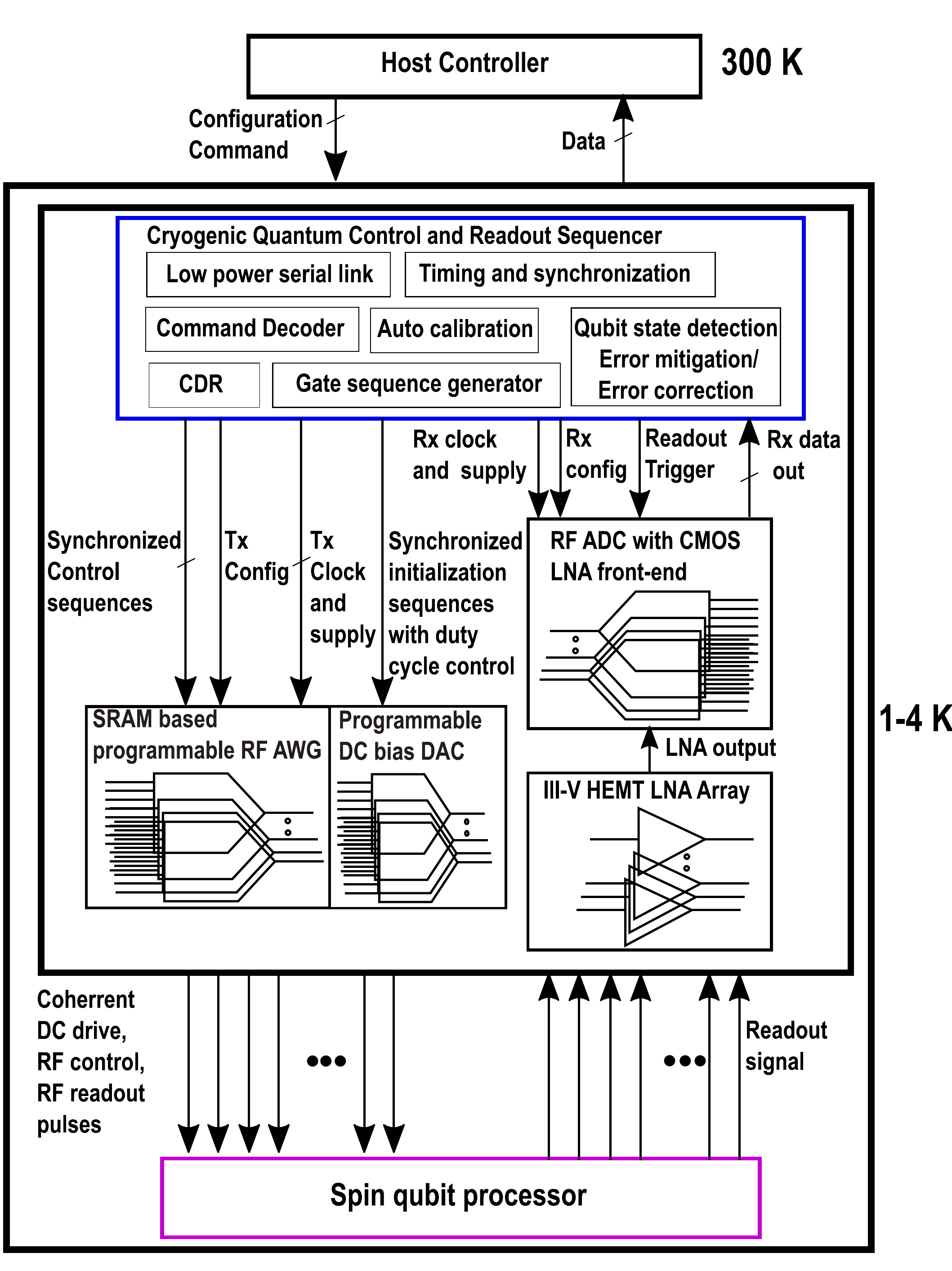}}
\hfill
\subfloat[Integrated readout chain \label{1c}]{%
	\includegraphics[width=0.3\linewidth]{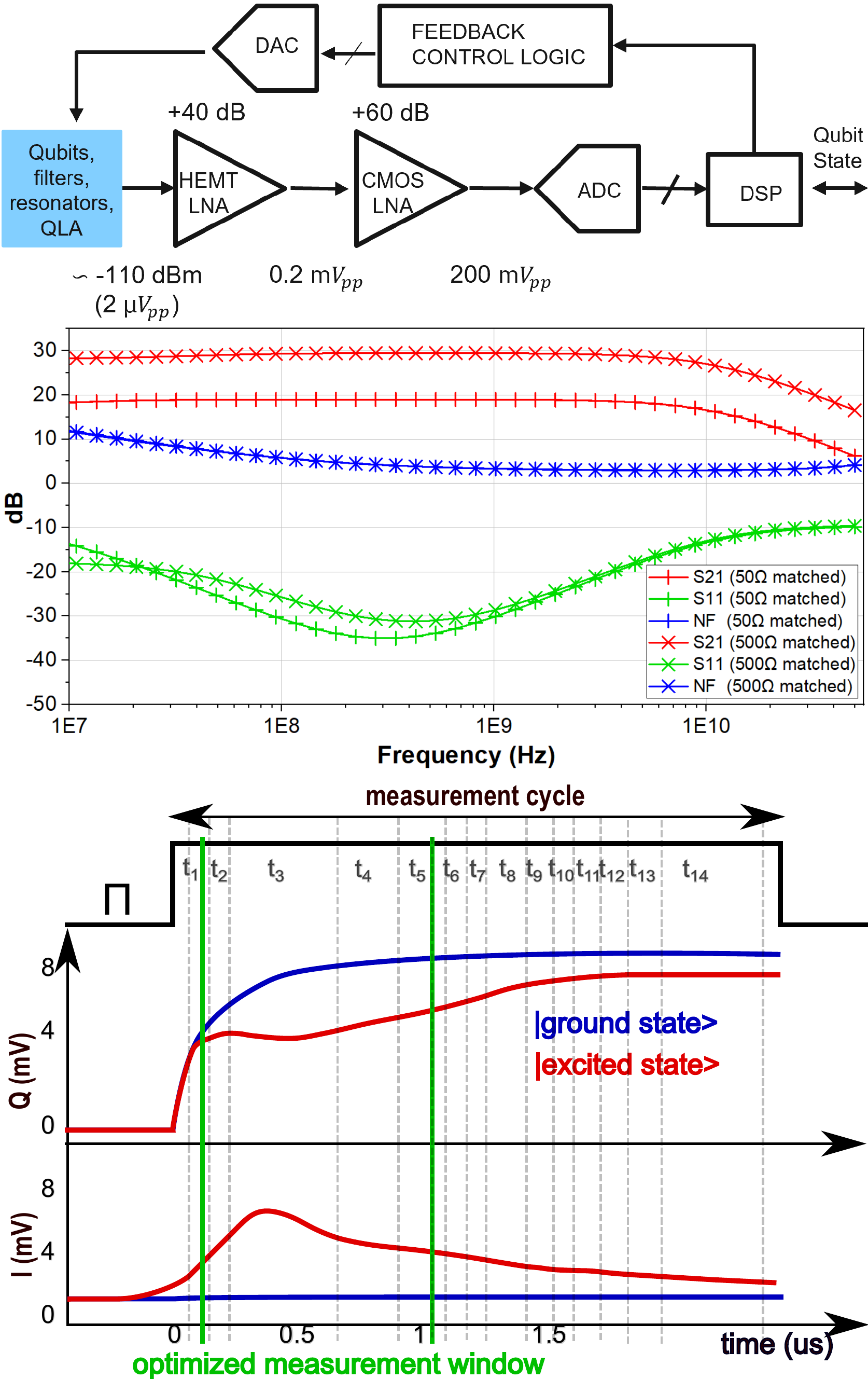}}
\hfill
\caption{(a) Architecture for error corrected quantum computing systems, showing the timing requirements. (b) Proposed integrated cryogenic control system co-packaged with hot spin qubits at 1-4\,K. (c) Block level schematic of an integrated feedback based qubit control chain (top), Simulated S-parameters and noise figure of CMOS LNA for the 50~$\Omega$ and 500~$\Omega$ matched design (middle), Example sketch of a typical qubit calibration curve showing division of time bins, used in a weighted multiply and accumulate operation at each time bins t\textsubscript{0} - t\textsubscript{14}, to discriminate the qubit state (bottom)}
\end{figure*}

The focus of this work is to analyze the system level requirements of control systems for large-scale quantum computers with 1000 physical qubits and above. The performance of a quantum computer depends upon several factors, such as the number of physical qubits, the number of gates that can be applied, the connectivity of the device, and the number of operations that can be run in parallel \cite{b4}. Realization of fault-tolerant quantum computers requires error correction by expanding the Hilbert space of logical qubits, which in turn requires coherent operation of many physical qubits \cite{b2}. The control and readout circuitry design, therefore, must explore ways to achieve dense integration and synchronized operation.
The proposed integrated control layer, \figurename\ref{1a} can deliver optimized control and readout signals, while reducing the cabling overhead in large scale quantum computing systems. It holds the potential for co-integration with high density qubit platforms such as spin systems (\figurename\ref{1b}), provided the thermal loading, noise and crosstalk criteria are satisfied. Power consumption has been, by far, the most important figure of merit concerning cryogenic CMOS circuit design.  We propose to shift the focus towards other important metrics such as feature size, low noise, synchronous operation, generation of low jitter clocks, and signal integrity. We introduce low power design strategies for cryogenic control systems, to facilitate signal processing within the cryostat and to reduce the data rate through serial links to room temperature, thus achieving low power through scaling offered by advanced CMOS nodes. 

A major challenge in the design of an integrated control and readout system is to minimize the decoherence caused due to coupling of the quantum processor to the external environment. It has been shown that thermal loading leads to qubit dephasing and reduced relaxation times \cite{b5}-\cite{b8}. Considerable effort has been made to isolate qubits from thermal radiation, by proper thermalization of the cables and other components such as attenuators and filters in the cryostat. To suppress thermal noise, filters are employed whose stop-bands lie well outside the frequency range of qubits and their readout circuitry. To reduce the infrared coupling, blocking filters have been proposed \cite{b7}-\cite{b8}. Despite these measures, other noise sources such as 1/f noise from the control electronics can affect the qubit coherence times. It has been shown that these effects can be mitigated by using optimized control signals \cite{b9}-\cite{b10}. 
\vspace{-5pt}
\section{Quantum control system architecture}
The current state of the art quantum control systems use rack mounted equipments or FPGA boards with off the shelf components to generate the required set of control and readout signals. The control signals are sent through attenuators to achieve the required levels of amplitude, of the order of tens of millivolts for spin qubits \cite{b15}. The qubit readout, on the other hand requires several amplification stages, including quantum limited amplifiers, HEMT LNAs and CMOS amplifiers to boost the signal amplitude from microvolts to hundreds of millivolts required for state discrimination. The main requirements of a generic quantum control system are: 
\begin{enumerate}
\item[1.] Generate arbitrary waveforms (eg: 2-20 GHz for spin qubits \cite{b13}) and 4-8 GHz for dispersive readout at the required amplitude, phase, pulse width with acceptable signal integrity, noise, and jitter. 
\item[2.] Amplify, digitize and process the readout signals from the quantum processor to determine the state of qubits and perform parity-based error correction
\item[3.] Generate reference clocks, bias voltages and currents for different blocks within the system, while adhering to the known specifications.  
\item[5.] A task scheduler to orchestrate the sequences applied to the qubit gates, the control and readout signals, the optimal position of the measurement window, recalibration and data transfer.
\item[6.] Fault detection mechanism 
\item[7.] Reliability with respect to environmental changes
\end{enumerate} 
\subsection{Timing and synchronization requirements}
The quantum control hardware puts forth hard real-time requirements on the synchronization among FPGA devices and digital logic with complex clocking schemes. The clock tree is designed with reference to the system clock in the order of 100 MHz. On board phase locked loops can be used to generate high frequency clocking for blocks such as  RF DAC (eg: 5 GHz reference) and ADC (eg: 500 MHz Nyquist rate). A ``sync" pulse must be propagated to each logical control unit shown in \figurename\ref{1a}  to ensure synchronous operation of waveform generators and receiver sampling. The timing block with counters generates and propagates synchronous trigger pulses to the scheduler finite state machine (FSM), that initiates the operation of control and readout blocks. Periodic monitoring of logical qubit controller is required to ensure that the system stays in sync mode. The timing diagram of a logical controller SoC is shown in \figurename\ref{2a}    

\subsection{Control chain}
\begin{figure*}[tbh]
	\centering
	\subfloat[Bock diagram of a qubit control and readout sequencer SoC with real time calibration and optimal measurement windowing logic \label{2a}]{%
		\includegraphics[width=0.48\linewidth,keepaspectratio]{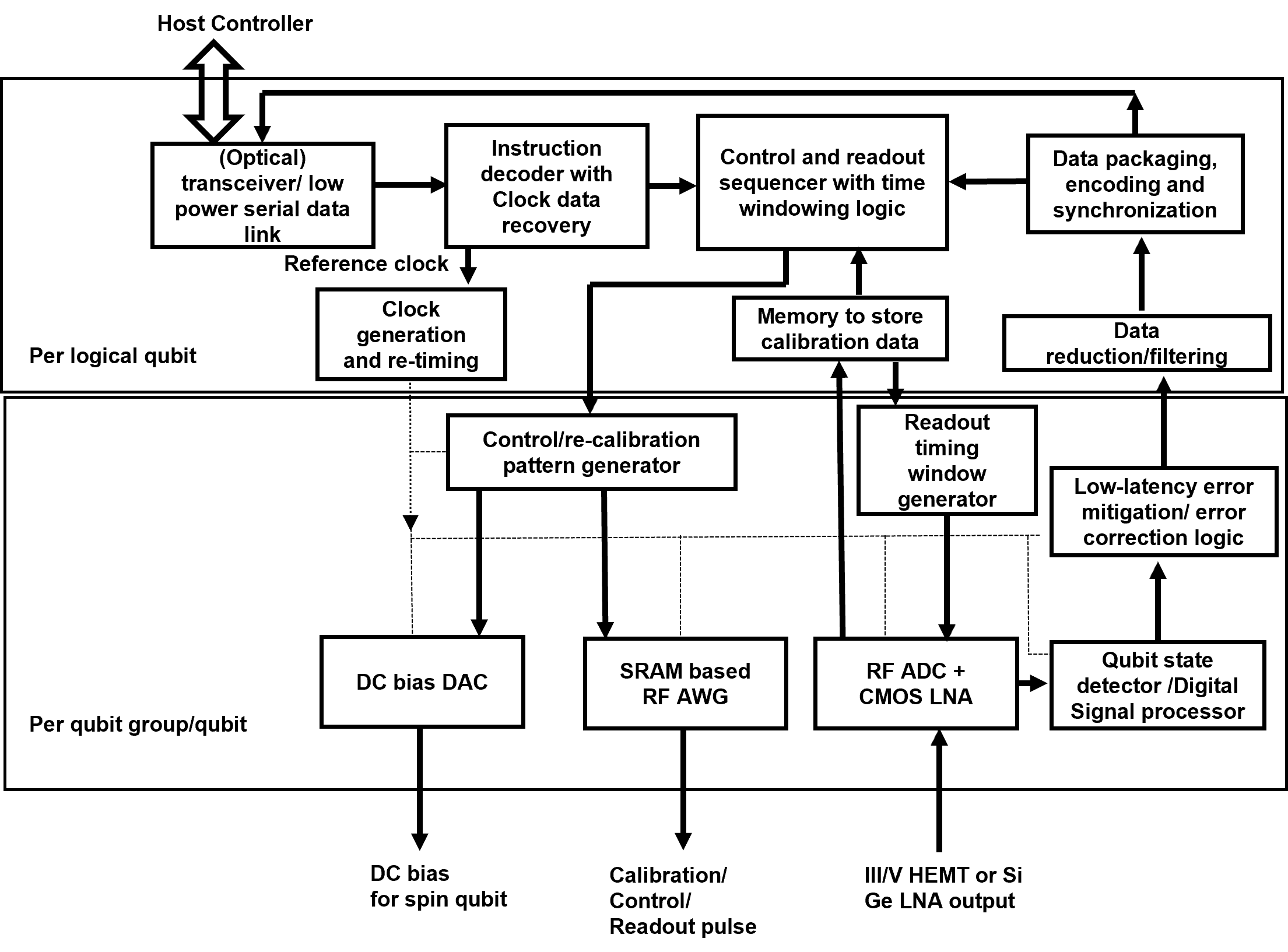}}
	\hfill
	\subfloat[Timing diagram of a logical qubit controller\label{2b}]{%
		\includegraphics[width=0.5\linewidth,keepaspectratio]{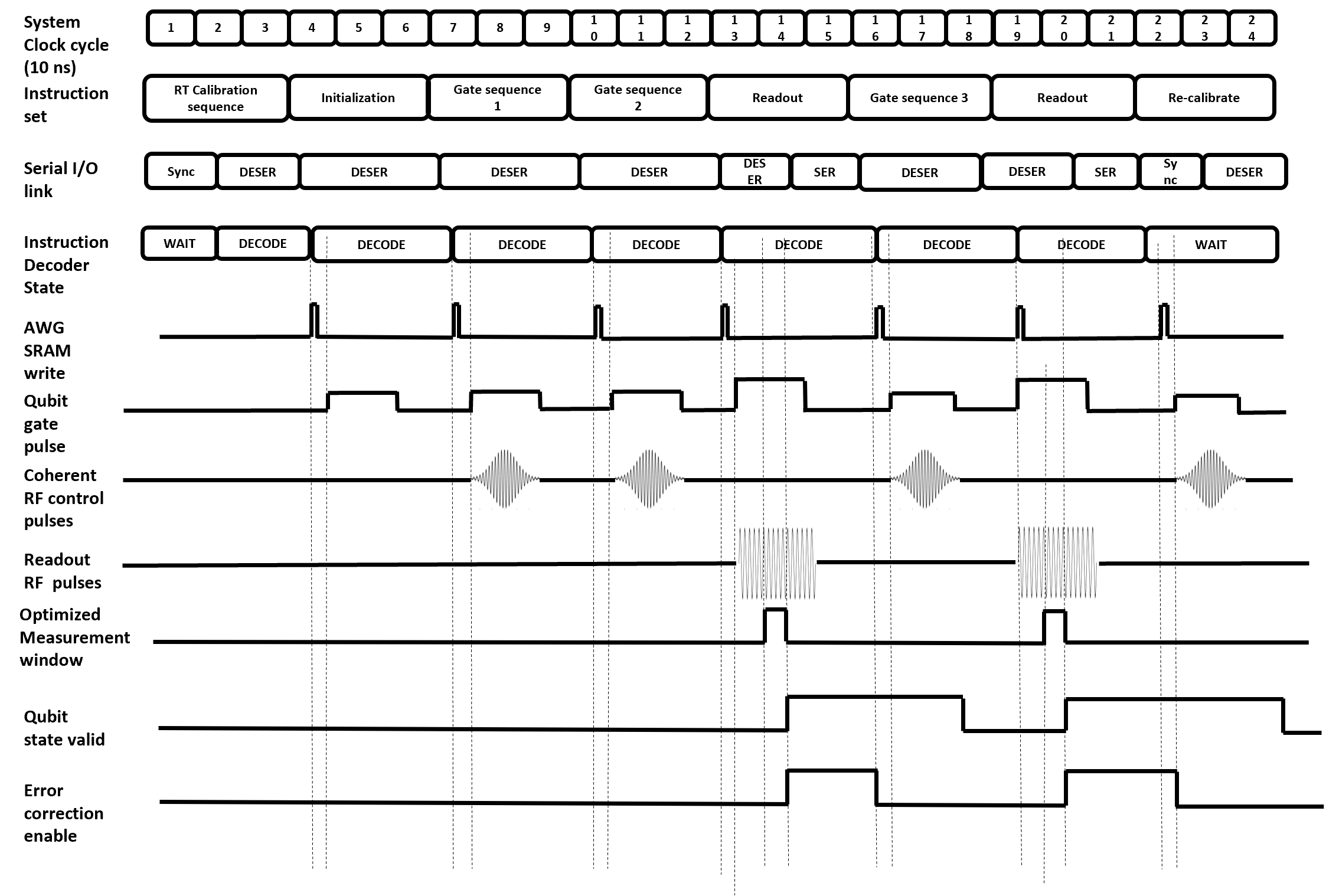}}
	\hfill
	\caption{(a) Block level representation of an integrated sequencer system on chip showing the hardware blocks required for deriving optimized timing window for measurement and control. (b) Timing diagram showing sequencer operation with synchronized trigger signals for control and readout, showing optimal window for readout and low latency error correction.}
	\label{fig2} 
\end{figure*}

The digital architecture of a sequencer system on chip \figurename\ref{2a} receives commands through serial links and decodes the instructions with parameters for gate control sequences. The control signal amplitude, pulse shape, phase and duration are stored in a waveform memory. Using direct digital synthesis, arbitrary waveforms are generated using a fixed frequency reference clock. An alternate approach for power reduction uses analog mixers and local oscillators (LO) to up convert the IF signals to GHz frequency range as required by the qubit hardware \cite{b11}-\cite{b14}. Since LO phase noise not only limits the signal to noise ratio but also demands additional phase rotation capabilities at the receiver, we propose to use direct digital synthesis approach using an RF DAC. Depending upon the frequency multiplexing ratio for control signals, the power budget per qubit can be calculated. 
\subsection{Readout chain}
Quantum computing systems have stringent requirements such as 99.9\% on readout fidelity of the qubit state. Our targeted spin systems employ dispersive readout schemes, owing to their minimal device overhead and synergy with super conducting systems. We propose a fully integrated readout chain with RF A/D converters as qubit digitizers with an integrated amplifier front-end and a digital signal processor back-end for qubit state discrimination as shown in \figurename\ref{1c}. Co-integration of CMOS low noise amplifiers with in-house developed InP HEMT LNA offer several benefits. From simulation shown in \figurename\ref{1c}, we infer that shifting the impedance matching to 500~$\Omega$ from 50~$\Omega$ results in 10 dB increase in the gain of an inverter-based CMOS LNA while yielding 10$\times$ reduction in power reported in table \ref{tableParams}. The simulated characteristics of single stage LNA is reported in table \ref{LNAcharacteristics}. The inverter based three stage CMOS LNA characteristics with 500~$\Omega$ input impedance provides the required gain of 60\,dB gain and a bandwidth of 5.8 GHz post layout in 5 nm FinFET technology. 
\begin{table}[tbh]
	\caption{Inverter with resistive feedback ($R_F$) single stage LNA design parameters in 5 nm FinFET technology}
	\label{tableParams}
	\centering
	\begin{tabular}{|c|c|c|c|c|}
		\hline
		\textbf{I/p impedance} & \textbf{$FET_1$ size} & \textbf{$FET_2$ size} & \textbf{$R_F$} & \textbf{Power}\\
		\hline
		50  $\Omega$ & 800 fins & 600 fins & 0.497 k$\Omega$ & 12.91 mW\\
		\hline
		500 $\Omega$ & 80 fins & 60 fins & 5.30 k$\Omega$ & 1.42 mW\\
		\hline
	\end{tabular}
\end{table}
\vspace{-10pt}
\begin{table}[tbh]
	\caption{Simulated single stage CMOS LNA characteristics}
	\begin{center}
		\centering
		\begin{tabular}{|c|c|c|c|}
			\hline
			\textbf{Bandwidth} & \textbf{Gain}& \textbf{Noise Figure}& \textbf{Power} \\
			\hline
			11.5 GHz & 29.5 dB & $<$3\,dB & 1.42 mW\\
			\hline
		\end{tabular}
		\label{LNAcharacteristics}
	\end{center}
\end{table}
\section{Strategies for power reduction} 
\subsection{Rapid power ON/OFF}
Depending upon the choice of error correction codes, the execution of gate sequences requires a subset of qubits and their respective control and readout circuitry to be active during an instruction cycle. We propose a technique that involves selective powering ON of the active circuitry in the qubit control and readout chain during the execution of quantum algorithms. Enabling the active nodes of the control and readout loop and powering OFF the redundant ones offers a huge advantage in terms of power efficiency and in turn makes room for  scalability. This can be achieved by generating synchronous power enable signals of control and readout blocks such as AWGs and A/D converters. It suppresses the total noise in the system, since the redundant blocks remain powered OFF during a given instruction cycle. This could potentially mitigate qubit dephasing induced by extensive operation of the control circuitry. A built-in rapid power ON capability achieves power saving when compared to the current state of the art systems that are powered ON throughout the entire control and measurement cycle. Furthermore, by reducing the ON time, CMOS circuits can be designed with more relaxed electromigration (EM) constraints and optimized with respect to additional parasitic capacitance, which would be required to fulfill the EM specs with longer power ON times. The risk of supply noise during power ON and how it affects the gate fidelity needs to be estimated. 

\subsection{Optimal measurement window from qubit calibration data}
Power consumption of CMOS circuits is directly proportional to their operation time. Reducing the operational time of the readout logic by using an optimized timing window derived from qubit calibration curve as illustrated in \figurename\ref{1c}, enables low power operation. The efficient measurement window for a given qubit group can be computed periodically by re-calibration as shown in \figurename\ref{2a} and \ref{2b}. It can be used to enable clock or power gating for digital and mixed signal blocks to achieve low power operation. 
A digitally synthesized qubit state discriminator based on weighted multiply and accumulate logic, consumes only \mbox{54.6 $\mu$W} of power with clock gating using optimized measurement window, as opposed to \mbox{256 $\mu$W} without clock gating.  

\subsection{Data reduction using low latency error correction and optimized feedback-based control}
Latency requirements on error correction hardware is expected to play a crucial role in large scale systems. The fidelity of fast qubit gates can be increased by optimized control pulses in a closed-loop fashion \cite{b9}-\cite{b10}. Low latency error correction logic at cryogenic temperature will help to generate optimized control signals using direct digital synthesis at GHz frequency range enabling faster gates. 
\subsection{Multiplexing}
Digital to analog converters with 2-20 GHz bandwidth and sufficient spurious free dynamic range can be used to generate control signals for multiple qubits in the current research environment \cite{b13}. The control of multiple qubits using frequency division multiple access (FDMA) has been proposed to reduce the cabling bottleneck due to scaling, however, this can be proven inefficient in some systems employing cross-resonant gates \cite{b14}. The real merits of using FDMA as opposed to 1:1 qubit to control circuit ratio for large scale error corrected systems needs in-depth evaluation. Frequency multiplexing may be favorable for dispersive readout, where RF signal bandwidth is within 4-8 GHz. 

%

\section*{Conclusions and outlook}
A study on the design of an integrated control system for quantum computers is reported. The proposed architecture uses digitally assisted RF AWG and RF ADC designs, with techniques to enable an integrated amplification chain and qubit state discriminator. The reported method eliminates the need of noisy local oscillators and analog mixers, while reaping the benefits of scaling in advanced CMOS nodes. It enables real time equalization and re-calibration to compensate drift due to changes in the cryostat cabling. Moreover, an optimized timing window can be derived to minimize the power consumption of readout circuitry. The proposed architecture supports cryogenic digital signal processing, leading to fully integrated, feedback-based optimized cryogenic control systems. In the near term, it is expected that the control hardware evolves with changing qubit hardware specifications. Therefore, it is desirable to have a high level of programmability at low cost and compactness. 

In the long-term, the features of subcomponents must be optimized, and their operating temperature must be defined in the interest of signal integrity, noise, and power consumption. Digital signal processing capabilities and state discrimination at cryogenic temperature is crucial to reduce the required data rate on serial links to room temperature. The versatility offered by the presented approach is desirable to accommodate evolving spin qubit specifications and to overcome the memory requirement, which is one of the biggest challenges of cryogenic CMOS systems. The discussed architecture also offers a solution for realizing large scale control systems with synchronized operation of multiple gates with inserted correction gates and the generation of their timing sequences.  
 
\section*{Acknowledgment}
This work was part of NCCR SPIN funded by the Swiss National Science Foundation (grant number 51NF40-180604).

\end{document}